\documentclass{article}

\PassOptionsToPackage{numbers, compress}{natbib}

\usepackage[final]{neurips_2021}

\usepackage[utf8]{inputenc} 
\usepackage[T1]{fontenc}    
\usepackage{hyperref}       
\usepackage{url}            
\usepackage{booktabs}       
\usepackage{amsfonts}       
\usepackage{nicefrac}       
\usepackage{microtype}      
\usepackage[table]{xcolor}         
\usepackage{graphicx}
\usepackage{amsmath,amsfonts,mathtools}
\usepackage{tabularx,longtable,multirow,caption,subcaption}

\DeclareMathOperator*{\argmin}{arg\,min}

\title{Cross Modality 3D Navigation Using Reinforcement Learning and Neural Style Transfer}

\author{%
    Cesare Magnetti$^1$\\
    \texttt{cm1320@ic.ac.uk}
    \And
    Hadrien Reynaud$^1$ \\
    \texttt{hjr119@imperial.ac.uk} \\
    \And
    Bernhard Kainz$^{1,2}$ \\
    \texttt{bkainz@imperial.ac.uk} \\

}

\begin{document}

\maketitle
\vspace*{-0.5cm}
\begin{center}
$^1$ Imperial College London, London, SW7 2BU, UK\\
$^2$ Friedrich-Alexander-Universität Erlangen-Nürnberg, 91054 Erlangen, DE  
\end{center}

\begin{abstract}
  This paper presents the use of Multi-Agent Reinforcement Learning (MARL) to perform navigation in 3D anatomical volumes from medical imaging. We utilize Neural Style Transfer to create synthetic Computed Tomography (CT) agent gym environments and assess the generalization capabilities of our agents to clinical CT volumes. Our framework does not require any labelled clinical data and integrates easily with several image translation techniques, enabling cross modality applications. Further, we solely condition our agents on 2D slices, breaking grounds for 3D guidance in much more difficult imaging modalities, such as ultrasound imaging. This is an important step towards user guidance during the acquisition of standardised diagnostic view planes, improving diagnostic consistency and facilitating better case comparison.
\end{abstract}

\section{Introduction}
\label{sec:introduction}
We propose a framework to automatically find the cardiac 4-chamber plane \cite{schoenhagen2005ct} in CT volumes. Currently, the analysis of these volumes is generally carried out by sampling images along the standard axes. Automatically sampling a view containing all relevant anatomical structures, would drastically speed up and improve the quality of the diagnosis. 
Code can be found at \url{https://github.com/CesareMagnetti/AutomaticUSnavigation}.

The current state-of-the-art in this domain often uses thick 3D planes sampled from manually labelled anisotropic volumetric scans as an input~\cite{alansary2018automatic,lu2011automatic}. Our method does not require labelled data and exploits the high-resolution segmentations of the XCAT virtual phantom~\citep{segars20104d} to guide the agent towards any selected anatomical structures of interest while generalising to real CT data.


\section{Methodology}
\label{sec:methods}

\begin{figure}[!htb]
     \centering
     \begin{subfigure}[b]{0.49\textwidth}
         \centering
         \includegraphics[width=\textwidth]{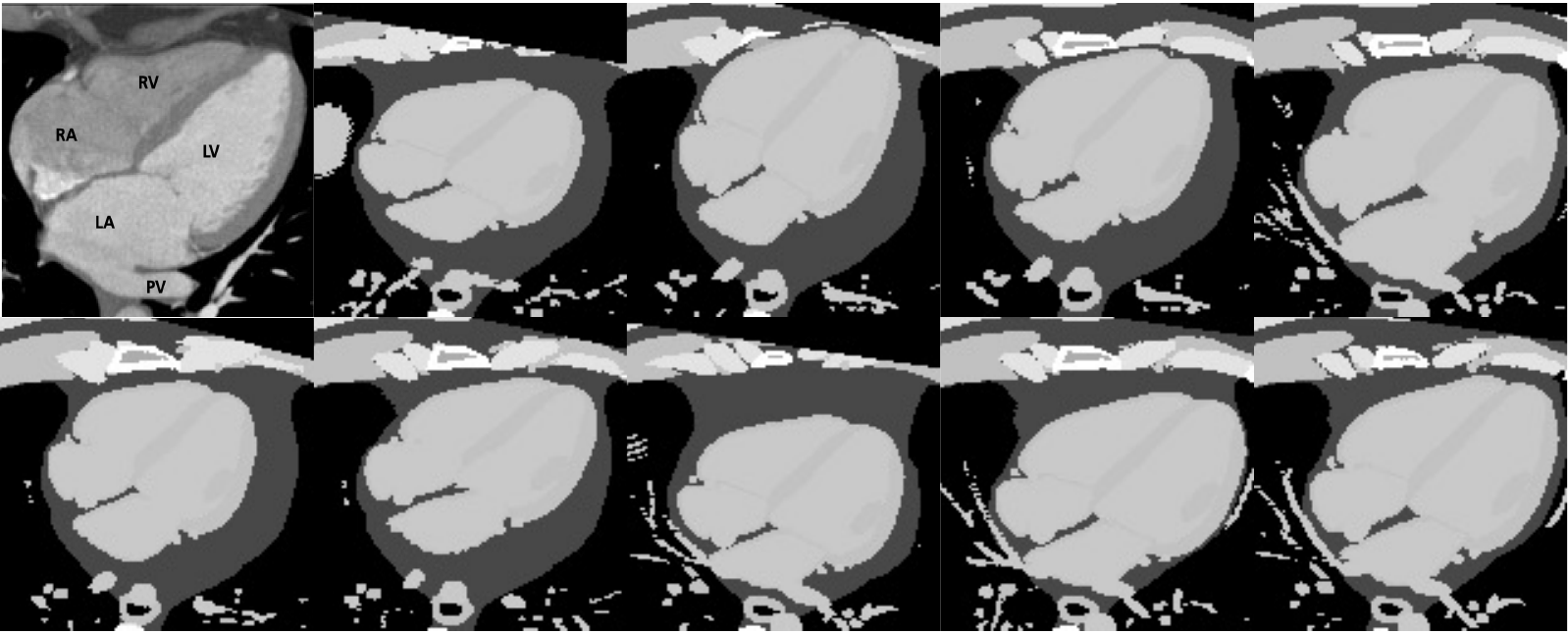}
         \caption{Approximated 4-Chamber planes XCAT}
         \label{fig:4ChamberViewsXCAT}
     \end{subfigure}
     \hfill
     \begin{subfigure}[b]{0.49\textwidth}
         \centering
         \includegraphics[width=\textwidth]{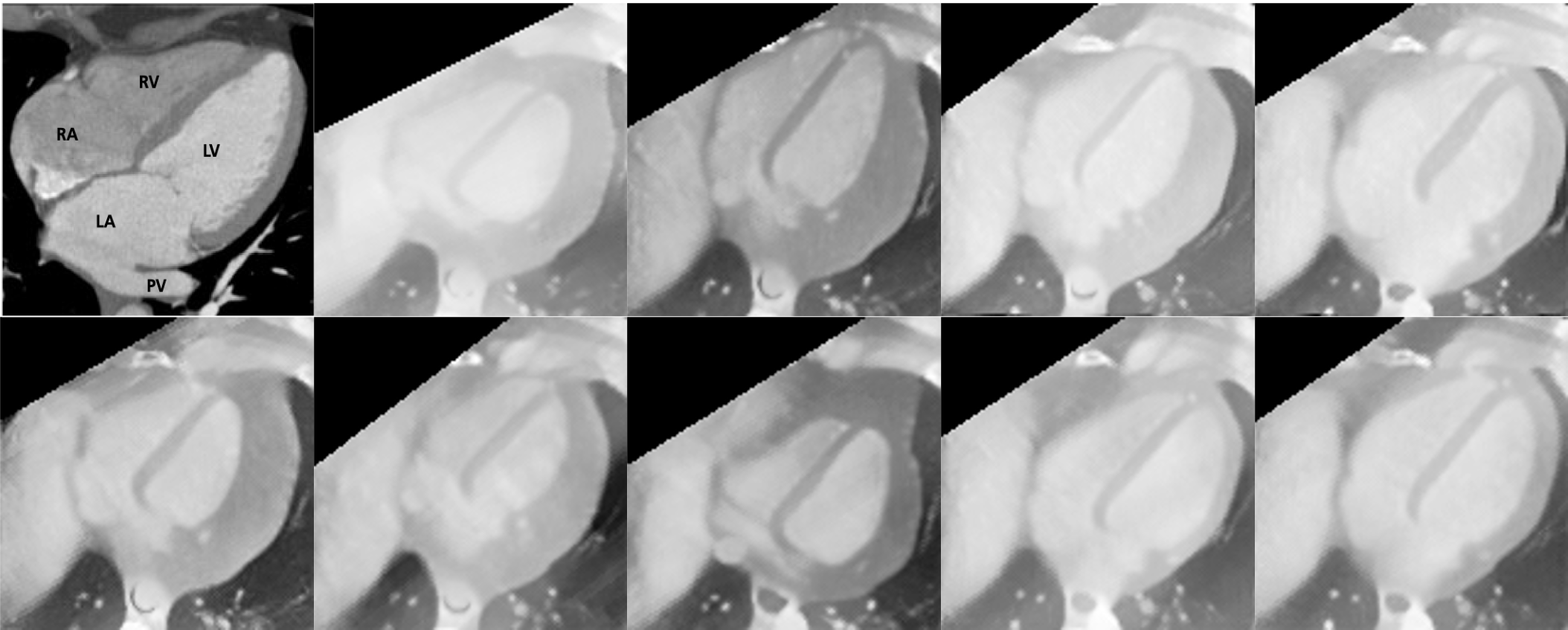}
         \caption{Approximated 4-Chamber planes fake CT}
         \label{fig:4ChamberViewsCT}
     \end{subfigure}
     \begin{subfigure}[b]{0.49\textwidth}
         \centering
         \includegraphics[width=\textwidth]{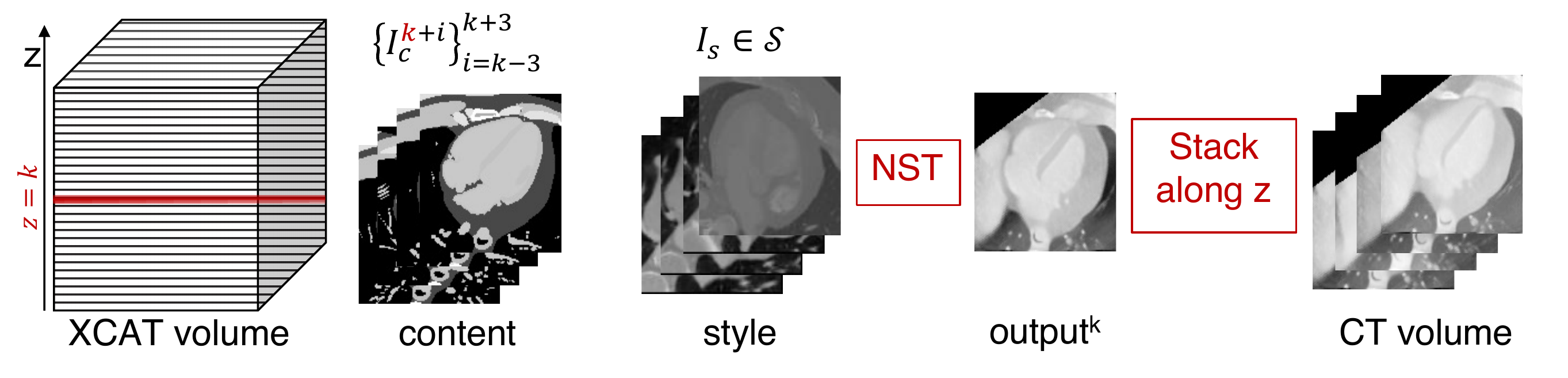}
         \caption{NST routine schematic}
         \label{fig:NSTroutine}
     \end{subfigure}
     \hfill
     \begin{subfigure}[b]{0.49\textwidth}
         \centering
         \includegraphics[width=\textwidth]{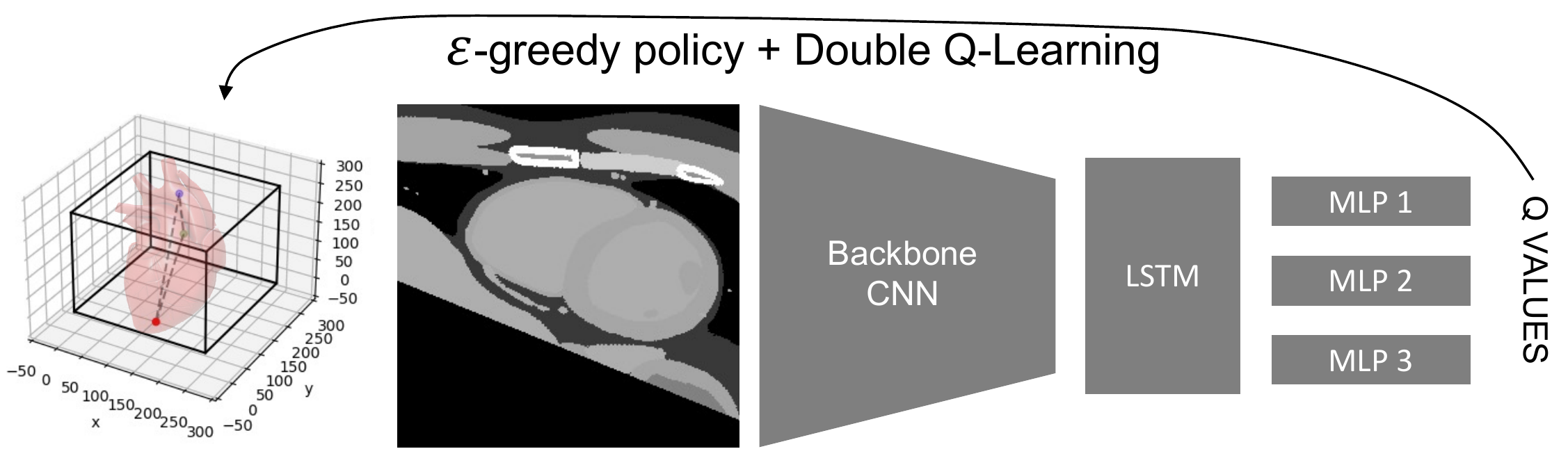}
         \caption{Navigation framework schematic}
         \label{fig:general_framework}
     \end{subfigure}
        \caption{a) Approximated 4-chamber planes on XCAT with reference taken from \cite{schoenhagen2005ct}. b) Approximated 4-chamber planes translated to have CT texture. c) Diagram of NST routine to translate XCAT volumes to synthetic CT volumes. d) General MARL navigation framework.}
\end{figure}

\textbf{Dataset and pre-processing} We use the XCAT program \cite{segars20104d} to generate 20 synthetic cardiac volumes of shape $128^3$ using 8-bit precision. We achieve some level of anatomical diversity by randomly adjusting the default hyperparameters within reasonable ranges ($\pm$ 10\% for most cases). This produces volumes with realistic spatial variety. We empirically observed that the XCAT model could only produce a limited amount of diversity, and therefore kept our volume count to 20. 
We automatically label the 4-Chamber plane as the plane that passes through the centroids of the left ventricle, right ventricle and left atrium. Note that we are able to find those centroids as XCAT comes with readily available semantic segmentations of the generated volumes. Some approximated 4-Chamber planes are shown in Figure \ref{fig:4ChamberViewsXCAT}.

We use Neural Style Transfer (NST) \cite{gatys2016image} to construct synthetic CT datasets, retaining content from the XCAT volumes and applying style from clinical CT scans taken from the LIDC-IDRI dataset \citep{armato2011lung}. Specifically, we slice each XCAT volume along the $z$ (axial) axis and retain style from a collection of 15 unique CT images per volume, and content from a window of $\pm 3$ slices along the $z$ axis. For more details, see Appendix \ref{app:NSTroutine}. Figures \ref{fig:4ChamberViewsCT}-\ref{fig:NSTroutine} respectively show a diagram of the NST routine and the translated 4-Chamber planes.

\textbf{General navigation framework}
Our MARL \citep{vlontzos2019multiple} framework consists of 3 agents free to move within a 3D volume. The action space of each agent has dimensionality $|\mathcal{A}| = 6$ with actions $a \in \{up, down, left, right, forward, backward\}$ corresponding to a 1-voxel-movement in the corresponding direction. Each configuration of the 3 agents will define a 2D plane ($a_0x+a_1y+a_2z+a_3=0$) within the 3D volume, for which we obtain plane coefficients $a_i$ starting from the agents coordinates. We normalize plane coefficients as $a_i = \nicefrac{a_i}{\sum_{i=0}^3 |a_i|}$, and we invert them if necessary, such that $a_0\geq0$. We then query a slice of anatomy taking the nearest neighbour projection of the volume voxels on that plane and pass it through a recurrent, multi-head DQN architecture \cite{mnih2015human,hausknecht2015deep} to predict the Q-values of each agents' movement. The whole architecture is then trained using an $e-greedy$ exploration policy coupled with Double Q-Learning \cite{van2016deep}. Fig. \ref{fig:general_framework} shows a diagram of the overall framework.

\textbf{Reward shaping} We take inspiration from \cite{alansary2018automatic} and reward our agents for decreasing the Euclidean distance between the plane coefficients of the current plane and the 4-Chamber plane. We additionally reward our agents for increasing the number of pixels in the current plane that belong to anatomical tissues of interest (namely the left ventricle, right ventricle and left atrium, but others can be added). In both cases we give a reward of $+1$ for improving upon the previous state, $-1$ for worsening and $0$ otherwise. We also added specific rewards to encourage good behaviours: 1) we maximize the area of the triangle spanned by the agents to prevent them from lining up, hindering the definition of a plane and 2) we penalize the agents when they move outside the volume boundaries to keep each discrete step significant enough to impact the plane. All rewards except the out-of-boundary penalty are shared amongst agents. Lastly, we empirically found that terminating an episode when the agents oscillate led to better performance than using an additional stopping action.

\section{Results}
\label{sec:results}

Table \ref{tab:results} shows quantitative evaluations of our agents trained on XCAT volumes and synthetic CT volumes. We evaluate the normalized plane distance and anatomical rewards on testing episodes, and report the mean reward.
We additionally report the final plane distance between terminal planes and the goal plane. Our results show that navigating in XCAT is generally easier than navigating in synthetic CT volumes, which is expected as we are adding significant noise to the input domain. 

\begin{table}[!htb]
\centering
\caption{Quantitative results of our agents on test volumes reported as $mean \pm std$ over 100 runs.}

\scriptsize{
\begin{tabular}{|c|c|c|c|c|}
\hline
\centering
\cellcolor{gray!20}Volume ID& \cellcolor{gray!20}data & \cellcolor{gray!20}plane distance reward $\uparrow$& \cellcolor{gray!20}anatomical reward $\uparrow$& \cellcolor{gray!20}distance from goal $\downarrow$\\
\hline
\centering
\multirow{2}{*}{1 (test)} & XCAT & $\boldsymbol{0.4471} \pm \boldsymbol{0.3383}$ & $0.2187 \pm 0.2060$ & $\boldsymbol{8.13\text{e-}7} \pm \boldsymbol{1.42\text{e-}6}$\\
& CT & $0.2481 \pm 0.4969$ & $\boldsymbol{0.3140} \pm \boldsymbol{0.2653}$ & $8.12\text{e-}6 \pm 5.81\text{e-}6$\\
\hline
\centering
 \multirow{2}{*}{2 (test)} & XCAT & $\boldsymbol{0.5106} \pm \boldsymbol{0.2946}$ & $0.1896 \pm 0.1684$ & $\boldsymbol{1.06\text{e-}6} \pm \boldsymbol{1.53\text{e-}6}$\\
 & CT & $0.3418 \pm 0.4819$ & $\boldsymbol{0.3139} \pm \boldsymbol{0.2608}$ & $7.91\text{e-}6 \pm 5.66\text{e-}6$\\
\hline
\centering
\multirow{2}{*}{3 (test)} & XCAT & $\boldsymbol{0.4240} \pm \boldsymbol{0.3387}$ & $0.1575 \pm 0.2071$ & $\boldsymbol{1.16\text{e-}6} \pm \boldsymbol{1.64\text{e-}6}$\\
& CT & $0.4172 \pm 0.4807$ & $\boldsymbol{0.2394} \pm \boldsymbol{0.2513}$ & $8.31\text{e-}6 \pm 6.04\text{e-}6$\\
\hline
\centering
\multirow{2}{*}{4 (test)} & XCAT & $\boldsymbol{0.4574} \pm \boldsymbol{0.3483}$ & $0.1712 \pm 0.2018$ & $\boldsymbol{1.02\text{e-}6} \pm \boldsymbol{1.65\text{e-}6}$\\
& CT & $0.4323 \pm 0.5004$ & $\boldsymbol{0.3264} \pm \boldsymbol{0.2277}$ & $9.41\text{e-}6 \pm 5.91\text{e-}6$\\
\hline
\centering
\multirow{2}{*}{5 (test)} & XCAT & $\boldsymbol{0.4499} \pm \boldsymbol{0.3535}$ & $0.1271 \pm 0.1935$ & $\boldsymbol{1.33\text{e-}6} \pm \boldsymbol{1.83\text{e-}6}$\\
& CT & $0.2934 \pm 0.5339$ & $\boldsymbol{0.3332} \pm \boldsymbol{0.2839}$ & $9.24\text{e-}6 \pm 6.75\text{e-}6$\\
\hline

\end{tabular}
}
\label{tab:results}
\end{table}

\begin{figure}[!htb]
     \centering
     \begin{subfigure}[b]{0.6\textwidth} 
         \centering
         \includegraphics[width=0.7\textwidth]{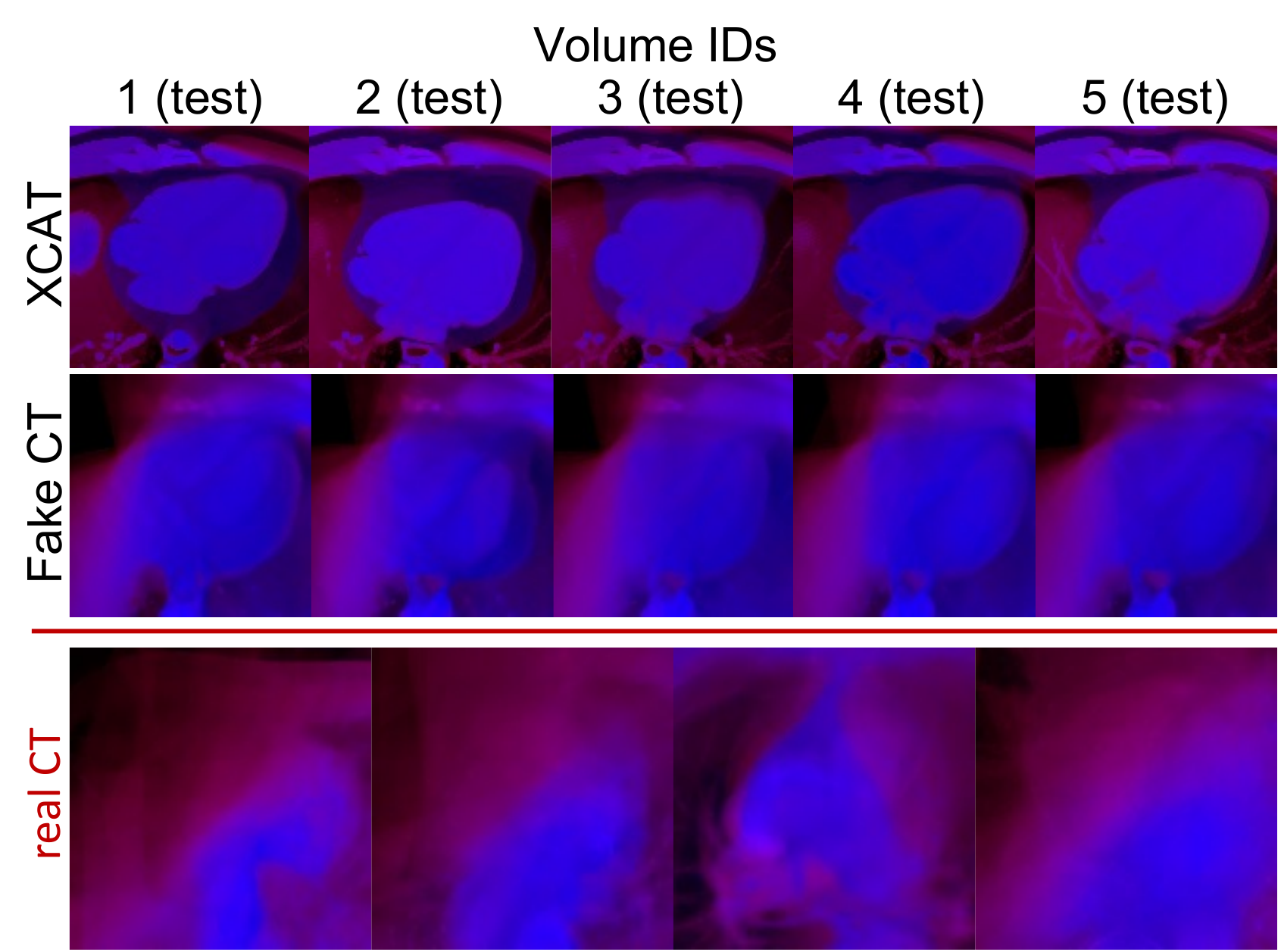}
         \caption{Qualitative results}
         \label{fig:resultsOverlays}
     \end{subfigure}
     \hfill
     \begin{subfigure}[b]{0.39\textwidth} 
         \centering
         \includegraphics[width=0.7\textwidth]{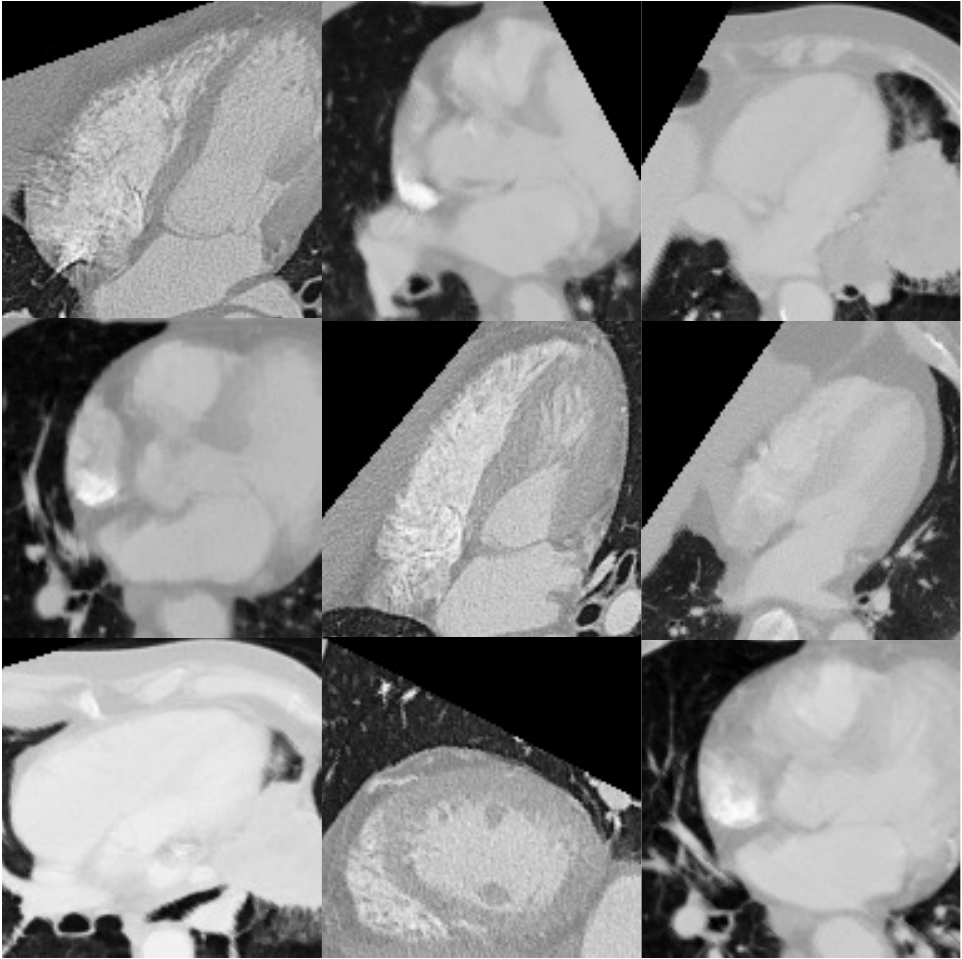}
         \caption{real CT terminal planes}
         \label{fig:realCTterminals}
     \end{subfigure}
        \caption{a) Mean (blue) and std (red) of pixel intensities over 100 terminal planes reached by our agents. b) Hand picked terminal planes on clinical CT scans.}
\end{figure}

Figure \ref{fig:resultsOverlays} shows the mean of the pixel intensities over $100$ terminal planes on the blue channel and the standard deviation on the red channel. Bright blue regions indicate consistency amongst our terminal planes and bright red regions indicate areas of high variation within our terminal planes. We report results for XCAT, fake CT and clinical CT scans from the LIDC-IDRI dataset \cite{armato2011lung}.  

The overlays show significant consistency in the terminal planes for both XCAT and synthetic CTs, showing precise 4-Chamber view averages. Overlays on real CTs show higher variance, with some confidence in retaining the left ventricle area of the image. This arises because real CTs are a much more noisy and complex input domain, and our data augmentation pipeline does not cover this whole input space yet. We believe this lack of diversity in our fake CTs to be the main limitation to the generalization of our model to most LIDC-IDRI CT volumes. Making our fake CTs more diverse should lead to a significant boost in our model performance on real CT scans. Figure \ref{fig:realCTterminals} shows hand-picked terminal planes of our agents deployed on clinical CT scans where we are able to get relatively close to a 4-chamber plane.

\section{Conclusion and Future Research}
\label{gen_inst}

Future work will focus on closing the reality gap between training our agents on simulated CT volumes and testing them on clinical CT scans. To this end we need to improve the NST routine to create more realistic and diverse synthetic CT volumes. This could be achieved by improving anatomical diversity of our XCAT volumes using elastic deformations, in order to generalize to variations in real patients' anatomy. We could also use 3D models to perform the NST routine rather than stacking 2D slices and homogenizing the plane intensities, as well as introduce some noise \& augmentations to improve texture. Another interesting route would be to register clinical CT volumes to fake CT volumes in order to obtain a paired dataset of segmented clinical CTs to train our agents.

We presented a novel framework for view planning towards 4-Chamber views in CT scans. Our method does not need any form of labelled clinical data thanks to its cross-modality training approach. 
Our Image-to-Image translation routine makes it straight forward to change the imaging domain of interest, by changing the translation process. Finally, our framework only conditions on 2D image inputs, therefore it can be extended to complex imaging modalities such as ultrasound, where it could enable sonographer guidance.

\section{Potential Negative Societal Impact}\label{sec:neg_soc}
To the best of our knowledge, we did not find any potential negative societal impact of our research. Regarding patient privacy, all clinical data used was anonymous and publicly available. Furthermore, our method would be applied for medical cases in the interest of patients and doctors, to speed up and standardize diagnosis. The AI does not conduct any direct diagnosis but rather aids doctors workflow.



\bibliographystyle{unsrt}
\bibliography{references}

\pagebreak
\appendix

\section{NST routine to create synthetic CT volumes}\label{app:NST}
\label{app:NSTroutine}
The overall optimization process at position $k\in [0, 128]$ along the $z$ axis is described in Equation \ref{eq:NSTroutine}.

\begin{equation}
    \begin{gathered}
    \mathcal{L}_{content}^k = \sum_{i=k-3}^{k+3}||F^l(I^k_t) - F^l(I^{k+i}_c)||^2_2\\
    \mathcal{L}_{style}^k = \sum_{I_s \in \mathcal{S}} \sum_l \frac{w_l}{4 H_l^2 W_l^2} ||F^l(I^k_t)F^l(I^k_t)^T - F^l(I_s)F^l(I_s)^T||_{F}^2\\
    \hat{I^k_t} = \argmin_{I^k_t} \mathcal{L}_{content} + \alpha \mathcal{L}_{style}\\
    \end{gathered}
    \label{eq:NSTroutine}
\end{equation}

Where $I_t^k$ is the target image at position $z=k$ (initialized as the XCAT slice $I_c^k$) and $F^l(\cdot)$ are the activations of layer $l$ of a VGG-19 \cite{simonyan2014very} pretrained on object recognition. Specifically, we retain style from layers $l = [conv_1, conv_2, conv_3, conv_4, conv_5]$, content from layers $l=[conv_4]$ and we set $\alpha=1e5$. For more details on NST refer to \cite{gatys2016image}. We translate each slice along the $z$ axis and we stack the final images to obtain our synthetic CT volume. We finally apply a first order Savitzky-Golay smoothing filter \cite{press1990savitzky} with a window of $5$ voxels to reduce misalignment artifacts generated when stacking all frames.

\section{Hyperparameters \& Resources}\label{app:HP}
Experiments are conducted on an Intel Xeon W-2123 CPU, with dual RTX Quadro 8000 and PyTorch 1.8.1+cu110. The training process is CPU heavy and lasts approximately 24 hours.

We train the Q-network for 2000 episodes of 125 steps parallelizing over 15 environment (training occurs every 15 time-steps), using learning rate $\lambda=0.0001$, discount factor $\gamma=0.999$ and exponentially decreasing $\epsilon = 1\rightarrow 0.005$. We use a batch size of 8 and a recurrent history length of 10, leading to an input batch of $15\times8\times10 = 1200$ frames to the backbone CNN for each train pass. We use soft updates \cite{kobayashi2021t} of the target network using a moving average of the Q-network weights ($\theta' = \tau \times \theta' + (1 - \tau) \times \theta$) with $\tau = 0.01$. For each environment we keep a prioritized replay buffer \cite{schaul2015prioritized} of capacity $C=25000$, prioritization $\alpha=0.6$ and exponentially increased bias correction $\beta=0.4 \rightarrow 1$. We end episode early if oscillations (visiting same states more than 3 times) are detected within a history of the past 20 visited states. When aggregating the rewards as explained in Section \ref{sec:methods}, we scale the area and out-of-boundary rewards by a factor of 0.01. Lastly, when we train on fake CTs, we randomize the window intensity of interest to embed the Q-network with generalization capabilities to various intensity levels of real CT scans.

\end{document}